\documentclass[aps,prl,twocolumn,amsmath,amssymb,showpacs,superscriptaddress,notitlepage,longbibliography]{revtex4-1}
\usepackage[colorlinks=true,linkcolor=blue,anchorcolor=red,citecolor=blue, urlcolor=blue]{hyperref}
\usepackage{bm}
\usepackage{graphicx}
\usepackage{color}

\begin{document}


\title{Anomalous thermoelectric effects of ZrTe$_{5}$ in and beyond the quantum limit}

\author{J. L. Zhang}
\affiliation{Anhui Province Key Laboratory of Condensed Matter Physics at Extreme Conditions, High Magnetic Field Laboratory of the Chinese Academy of Sciences, Hefei 230031, Anhui, China}

\author{C. M. Wang}
\email{Corresponding authors: wangcm@shnu.edu.cn}
\affiliation{Department of Physics, Shanghai Normal University, Shanghai 200234, China}
\affiliation{Peng Cheng Laboratory, Shenzhen 518055, China}
\affiliation{Institute for Quantum Science and Engineering and Department of Physics,
Southern University of Science and Technology, Shenzhen 518055, China}

\author{C. Y. Guo}
\affiliation{Institute of Material Science and Engineering, \'{E}cole Polytechnique F\'{e}d\'{e}ral de Lausanne (EPFL), 1015 Lausanne, Switzerland}

\author{X. D. Zhu}
\affiliation{Anhui Province Key Laboratory of Condensed Matter Physics at Extreme Conditions, High Magnetic Field Laboratory of the Chinese Academy of Sciences, Hefei 230031, Anhui, China}

\author{J. Y. Yang}
\affiliation{Anhui Province Key Laboratory of Condensed Matter Physics at Extreme Conditions, High Magnetic Field Laboratory of the Chinese Academy of Sciences, Hefei 230031, Anhui, China}

\author{Y. Q. Wang }
\affiliation{Anhui Province Key Laboratory of Condensed Matter Physics at Extreme Conditions, High Magnetic Field Laboratory of the Chinese Academy of Sciences, Hefei 230031, Anhui, China}

\author{Z. Qu }
\affiliation{Anhui Province Key Laboratory of Condensed Matter Physics at Extreme Conditions, High Magnetic Field Laboratory of the Chinese Academy of Sciences, Hefei 230031, Anhui, China}

\author{L. Pi}
\affiliation{Anhui Province Key Laboratory of Condensed Matter Physics at Extreme Conditions, High Magnetic Field Laboratory of the Chinese Academy of Sciences, Hefei 230031, Anhui, China}

\author{Hai-Zhou Lu}
\email{luhaizhou@gmail.com}
\affiliation{Institute for Quantum Science and Engineering and Department of Physics,
Southern University of Science and Technology, Shenzhen 518055, China}
\affiliation{Peng Cheng Laboratory, Shenzhen 518055, China}
\affiliation{Shenzhen Key Laboratory of Quantum Science and Engineering, Shenzhen 518055, China}

\author{M. L. Tian}
\email{tianml@hmfl.ac.cn}
\affiliation{Anhui Province Key Laboratory of Condensed Matter Physics at Extreme Conditions, High Magnetic Field Laboratory of the Chinese Academy of Sciences, Hefei 230031, Anhui, China}
\affiliation{Collaborative Innovation Center of Advanced Microstructures, Nanjing University, Nanjing 210093, China}

\date{\today}

\begin{abstract}

Thermoelectric effects are more sensitive and promising probes to topological properties of emergent materials, but much less addressed compared to other physical properties. Zirconium pentatelluride (ZrTe$_{5}$) has inspired active investigations recently because of its multiple topological nature.
We study the thermoelectric effects of ZrTe$_{5}$ in a magnetic field and find several anomalous behaviors. The Nernst response has a steplike profile near zero field when the charge carriers are electrons only, suggesting the anomalous Nernst effect arising from a nontrivial profile of Berry curvature. Both the thermopower and Nernst signal exhibit exotic peaks in the strong-field quantum limit. At higher magnetic fields, the Nernst signal has a sign reversal at a critical field where the thermopower approaches to zero. We propose that these anomalous behaviors can be attributed to the Landau index inversion, which is resulted from the competition of the $\sqrt{B}$ dependence of the Dirac-type Landau bands and linear-$B$ dependence of the Zeeman energy ($B$ is the magnetic field). Our understanding to the anomalous thermoelectric properties in ZrTe$_{5}$ opens a new avenue for exploring Dirac physics in topological materials.

\end{abstract}


\maketitle

{\color{blue}\emph{Introduction}}--
Transition-metal pentatellurides (ZrTe$_{5}$, HfTe$_5$...) have attracted considerable interest as topological materials very close to the boundary of topological phase transition \cite{Weng}. The negative longitudinal magnetoresistance \cite{LiQ,ZhengGL} and the linear energy-momentum dispersion demonstrated by the magnetoinfrared and optical spectroscopy measurements \cite{ChenRYPRB,ChenRYPRL,YuanX,ChenZG} implied a Dirac semimetal phase of ZrTe$_{5}$. In contrast, scanning tunneling spectroscopy and angle-resolved photoemission spectroscopy measurements detected a bulk band gap with topological edge states at side surfaces, giving the signatures of a weak 3D topological insulator (TI) \cite{WuR,LiXB,XiongH}. Other spectroscopic studies favor the strong TI phase \cite{Manzoni2017,Manzoni}. Recently, it was further proposed that these
states can be tuned by temperature or pressure \cite{Manzoni,XuB,ZhangJL}. In addition to its nature of multiple topological phases, a moderate magnetic field is enough to drive this layered material into the quantum limit, in which all carriers occupy the lowest Landau band.
This provides a platform to explore the exotic quantum phenomena caused by unique band topology in extremely strong magnetic fields. In particular, the magnetoresistance of ZrTe$_{5}$ decreases drastically when the field exceeds 8 T. Based on the picture of massless Dirac fermions, the sudden drop of magnetoresistance was conjectured to originate either from dynamical mass generation or topological phase transition from a 3D Weyl semimetal to a 2D massive Dirac metal \cite{LiuYW,ZhengGL2017}. Very recently, in ZrTe$_{5}$ the 3D quantum Hall effect was observed, which collapses into an exotic insulating state in the quantum limit \cite{TangFD}.

In the presence of a perpendicular magnetic field and a longitudinal thermal gradient, the diffusion of carriers can produce a longitudinal electric field $E_{x}=-S_{xx}\cdot|\nabla T|$ (thermopower) and a transverse electric field $E_{y}=S_{xy}\cdot|\nabla T|$ (the Nernst effect) \cite{BehniaK}. Since these thermoelectric effects are proportional to the derivative of the conductivities, they are more sensitive to anomalous contributions and have been used to study topological materials \cite{FauqueB,LiangTNC,ZhuZ,JiaZZ,GoothJ,MatusiakM,WatzmanSJ}. Zirconium pentatelluride, as a thermoelectric material, has been known for its large thermopower for nearly four decades \cite{Jones}. Nevertheless, the studies focusing on the thermoelectric properties of ZrTe$_{5}$ in magnetic fields are rare, especially in the quantum limit.

In this Letter, we study the thermopower and Nernst effect of ZrTe$_{5}$ single crystals. At low temperatures, the behavior of the Nernst signal is characterized by a step-like profile near zero field, suggesting the anomalous Nernst effect (ANE) resulted from the Berry curvature. By fitting with an empirical formula, we find that the anomalous component decreases with increasing temperature and becomes negligible above 30 K. Moreover, when the magnetic field has driven the system in the quantum limit, both the thermopower and Nernst signals present a broad peak which is distinct from quantum oscillations. Intriguingly, further increasing the magnetic field, there is a sign change in $S_{xy}$ at a critical field $B^*$ where $-S_{xx}$ converges to zero. Detailed theoretical analysis shows that the anomalous behaviors can be explained by the Landau index inversion, which originates from the 3D Dirac fermions in ZrTe$_5$.

\begin{figure}[tb]
\centering
\includegraphics[width=\columnwidth]{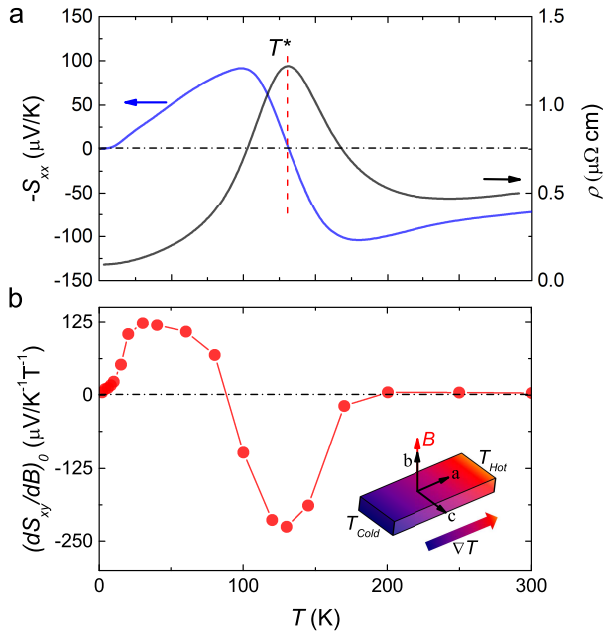}
\caption{(a) Temperature dependence of
the electrical resistivity $\rho(T)$ (black) and Seebeck coefficient -$S_{xx}(T)$ (blue) of ZrTe$_{5}$ at zero magnetic field. (b) Temperature dependence of the derivative of the Nernst signal with respect to the magnetic field at zero field $(dS_{xy}/dB)_0$. Inset: the measurement setup. $B$ is the magnetic field and $\nabla T$ is the temperature gradient. $a$, $b$, and $c$ are crystallographic axes.
}\label{fig.1}
\end{figure}

High quality single crystals of ZrTe$_{5}$ were synthesized using the iodine vapor transport method in a two-zone furnace \cite{Kam}. $-S_{xx}$ and $S_{xy}$ was measured with a standard one-heater-two-thermometers setup in a He$^{4}$ cryostat from 1.8 K to 300 K. A thermal gradient $\nabla T$ was applied along the $a$-axis. The voltage contacts were made with spot welding and each contact resistance was better than 1$\Omega$. In order to exclude the antisymmetric effect in magnetic fields, $-S_{xx}$ and $S_{xy}$ are symmetrized in positive and negative magnetic fields. The high-field measurements up to 33 T were performed in Chinese High Magnetic Field Laboratory at Hefei using a resistive water-cooled magnet. In the high-field setting, we measured the variation of the voltage produced by a constant heat flow as a function of the magnetic field. The absolute magnitudes of $-S_{xx}$ and $S_{xy}$ were calibrated later in another superconducting magnet (see the calibration details in Ref. \cite{SI}).

{\color{blue}\emph{Zero-field thermoelectric properties of ZrTe$_5$}}--
Figure 1, presents the temperature dependence of the thermopower $-S_{xx}$ and the derivative of the Nernst signal with respect to the magnetic field at zero magnetic field $(dS_{xy}/dB)_{0}$. At high temperatures ($>$ 132 K), a negative $-S_{xx}$ reveals that the dominant carriers are holes. As temperature decreases, $-S_{xx}$ shifts to positive at around $T^{*}$ = 132 K, where the resistivity shows a peak, indicating that ZrTe$_{5}$ evolves from a $p$-type semiconductor to $n$-type semimetal, consistent with the previous studies \cite{ZhangY,XuB}. In contrast to the Hall coefficient, which exhibits a sign reverse near $T^{*}$, the sign of $(dS_{xy}/dB)_{0}$ remains unchanged until temperature drops below 90 K.
At low temperatures ($<$ 90 K), both the positive $-S_{xx}$ and $(dS_{xy}/dB)_{0}$ show that the carriers are electrons only.

\begin{figure*}[tb]\centering
\includegraphics[width=18cm]{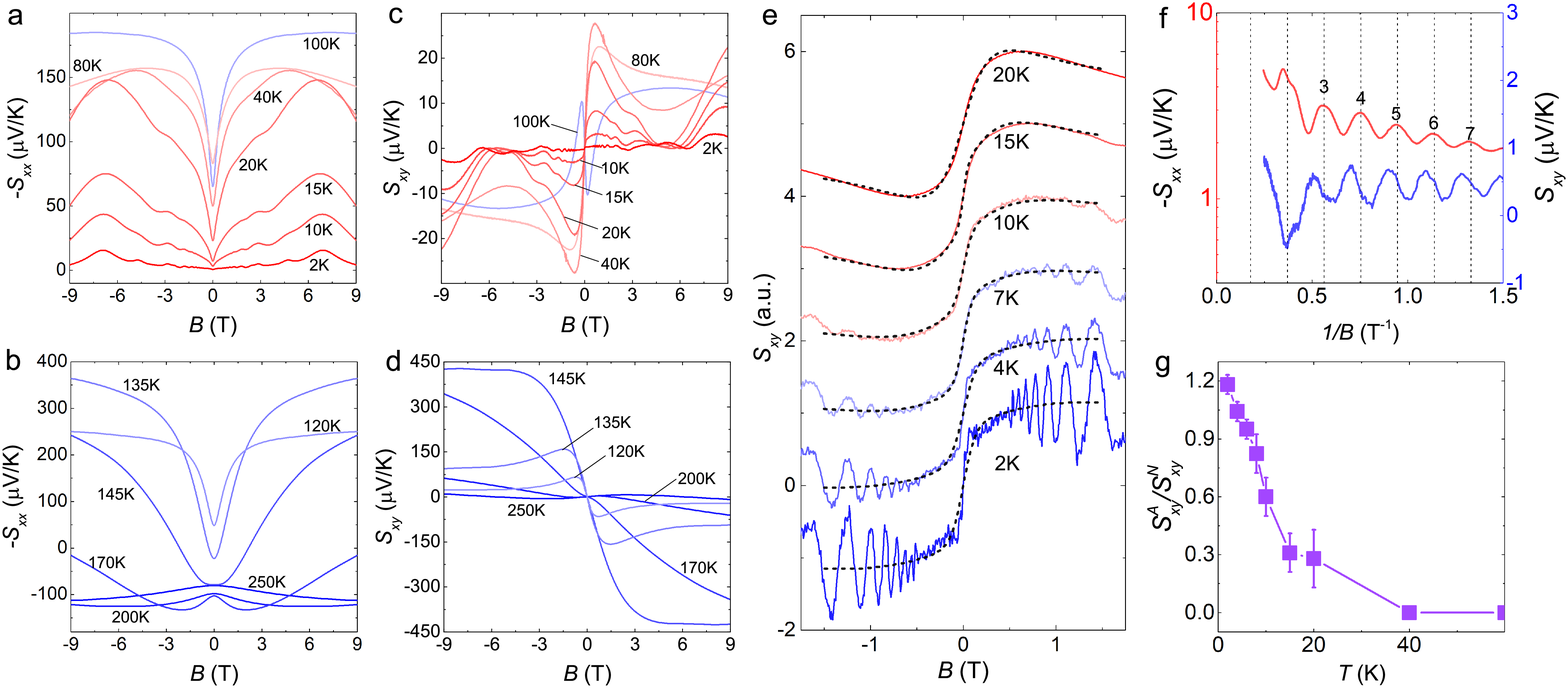}
\caption{Magnetic field dependence of the thermopower $-S_{xx}$ [(a)-(b)] and Nernst signal $S_{xy}$ [(c)-(d)] from 1.8 K to 300 K. (e) $S_{xy}$ at low fields at several temperatures below 20 K. The data is normalized and shifted for clarity. The black dash lines represent the fitting with Eq. (1). (f) $-S_{xx}$ and $S_{xy}$ as functions of $1/B$ at 1.8 K. The integer indices are marked at the maxima of $-S_{xx}$. (g) The temperature dependence of the ratio between the anomalous and conventional Nernst signals $S_{xy}^{A}/S_{xy}^{N}$, fitted by using Eq. (1).}\label{fig.2}
\end{figure*}

{\color{blue}\emph{Thermopower and Nernst signal at low magnetic fields}}--
Now we turn to investigate the field dependence of $-S_{xx}$ and $S_{xy}$. As shown in the inset of Fig. 1b, the magnetic field is applied along the $b$-axis, perpendicular to the thermal gradient $\nabla T$. At temperatures above the $p-n$ transition, although the dominant carriers are holes, both $-S_{xx}(B)$ and $S_{xy}(B)$ show complex behaviors.
At temperatures between 30 K and 130 K, $-S_{xx}(B)$ displays a saturation at low fields whose profile could be well fitted by the single-band Boltzmann-Drude model \cite{SI}. $S_{xy}(B)$, on the other hand, manifests a sharp Drude-like peak only at temperatures below 90 K, where the thermally activated hole carriers are undetectable. As temperature drops below 30 K, $-S_{xx}(B)$ starts to deviate from the single-band model fitting. In particular, below 10 K, $-S_{xx}(B)$ even grows quasi-linearly with increasing magnetic field up to 5 T \cite{SI}. The oscillations of $-S_{xx}$ and $S_{xy}$ are strong compared to the background. By employing the fast Fourier transform we obtain only one frequency  $F=$ 5.2 T. As shown in Fig. 2f, the integer Landau indices can be clearly identified from the peaks positions of $-S_{xx}$ ($S_{xy}$ is the off-diagonal term of the tensor, its extrema are shifted by 1/4 period compared to $-S_{xx}$). The linear extrapolation of the Landau indices intercepts at 0 $\pm$ 0.01, indicating a non-trivial $\pi$ Berry phase \cite{WangCM}. Besides, $(dS_{xy}/dB)_0$ has a maximum at 30 K, then decreases quadratically with decreasing temperature. All results suggest that another type of electron carriers, which are from the non-trivial bands, begin to dominate the magnetotransport properties below 30 K, leading to the quasi-linear field dependence of the thermopower.

At lower temperatures, as shown in Fig. 2e, $S_{xy}$ climbs to a maximum value at weak magnetic fields and tends to saturate. Although there are two types of electron carriers at low temperatures, their mobility and carrier density saturate below 20 K and the conductance ratio remains almost constant \cite{LiuYW}, thus the low-field $S_{xy}(B)$ is expected to be temperature independent. For this reason, the multiple carriers are unlikely to explain the step-like profile. On the other hand, the anomalous Hall effect has been claimed in the $p$-type ZrTe$_{5}$ ($T^{*}$ = 5 K) \cite{LiangTNP}. Despite the different carrier types in that and our ZrTe$_{5}$, the band structure is expected to be intrinsically the same \cite{ShahiP}. In this sense, the step-like $S_{xy}$ could be regarded as a signature of the anomalous Nernst effect. As discussed above, ZrTe$_{5}$ hosts both trivial and non-trivial electrons at low temperatures. In order to distinguish the conventional ($N$) and anomalous ($A$) Nernst signals, we fit the low-field data by using an empirical formula
\begin{equation}
S_{xy}^{tot}=S_{xy}^{A}\tanh(\frac{B}{B_{0}})+S_{xy}^{N}\frac{\mu B}{1+(\mu B)^{2}},
\end{equation}
where $\mu$ is the carrier mobility and $B_{0}$ is the saturation field above which the signal reaches its plateau value \cite{LiangT}. $S_{xy}^{N}$ and $S_{xy}^{A}$ are the amplitudes of the conventional and anomalous Nernst signals, respectively. As shown in Fig. 2e, the experimental data are well fitted by the empirical formula. At 2 K, the magnitudes of the anomalous and conventional Nernst signals are comparable. As temperature increases, $S_{xy}^{A}/S_{xy}^{N}$ drastically decreases (Fig. 2g), similar  to the anomalous Nernst effect of the topological Dirac semimetal Cd$_{3}$As$_{2}$ \cite{LiangT}. Different from Cd$_{3}$As$_{2}$, ZrTe$_5$ has a band gap at the $\Gamma$ point.

{\color{blue}\emph{Anomalous thermopower and Nernst signal in the quantum limit}}--
According to Fig. 2f, for magnetic fields above 5.2 T, all electrons should occupy the lowest Landau band, i.e., the system is in the quantum limit. In the quantum limit, the thermopower of a Dirac/Weyl semimetal is expected to grow linearly and non-saturatingly with increasing magnetic field \cite{SkinnerB}. However, in our cases, $-S_{xx}$ exhibits an unexpected board peak above 5 T. In order to further elucidate the unusual thermoelectric response in the quantum limit, we increase the magnetic field up to 33 T. As shown in Fig. 3, $-S_{xx}$ starts to drop around 7 T, then reaches a minimum around 15 T. Further increasing the magnetic field up to 33 T, $-S_{xx}$ turns to increase. Correspondingly, a hump emerges in the Nernst signal $S_{xy}$ right after the system enters the quantum limit. At a first glance, these anomalous features look like part of the quantum oscillation from the Landau bands (0,$+$) or (0,$-$). However, the amplitude of the peak is much higher than those of the quantum oscillations.

Another anomalous feature is that the Nernst signal $S_{xy}$ changes its sign at around $B^{\ast} = $ 14 T, where the thermopower $-S_{xx}$ at different temperatures converge to zero. The change of carrier type could lead to a sign reversal in the Nernst/Hall signal. Perviously, a field-induced sign change of $\rho _{yx}$ was observed in the Weyl semimetal TaP, in which the lowest Landau band move above the chemical potential in an extremely strong magnetic field, leading to a dramatic reduction of the carriers in the Weyl electron pockets \cite{ZhangCL}. As we discussed above, in our samples, the charge carriers at low temperatures are electrons only, and there is no hole-like band near the Fermi level \cite{ZhangY}. Moreover, the Hall resistivity $\rho_{yx}$ varies smoothly near $B^{\ast}$ \cite{SI}. Thus, the change of carrier type is unlikely to explain the anomalous sign reversal of $S_{xy}$.

\begin{figure}[htbp]
\centering
\includegraphics[width=8cm]{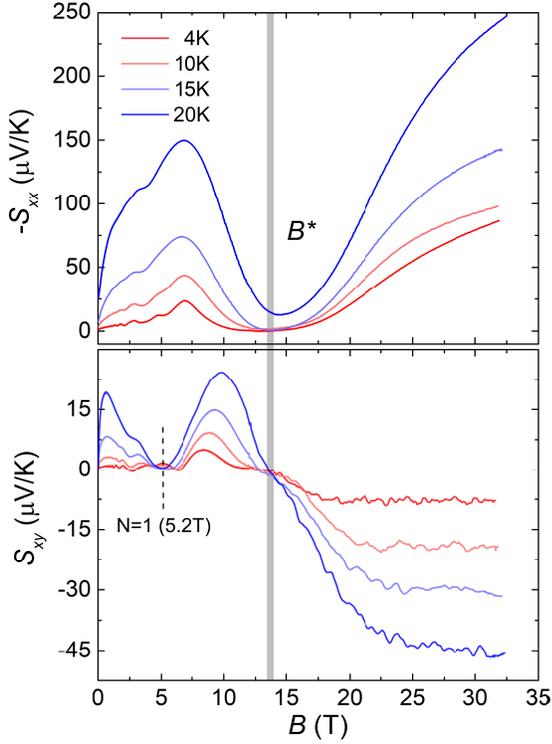}
\caption{(a) High-field measurements of $-S_{xx}$ up to 33 T at several temperatures. (b) High-field measurements of $S_{xy}$ at several temperatures. The critical field $B^*$ is indicated by the gray line around 14 T where $S_{xy}$ changes its sign and $-S_{xx}$ converges to zero.}\label{fig.3}
\end{figure}

{\color{blue}\emph{Landau index inversion beyond the quantum limit}}--
We now explore the underlying mechanism for the anomalous $-S_{xx}$ and $S_{xy}$ in the quantum limit. We use a 3D massive Dirac Hamiltonian, which was derived from the low-energy effective $k\cdot p$ Hamiltonian in the presence of the spin-orbit coupling \cite{ChenRYPRL}.
The Landau bands for index $\nu\ge1$ are written as
\begin{align}
  E_{\nu s\lambda}
  =&s\sqrt{\left[\sqrt{\nu \frac{2eBv^2}{\hbar}+m^2}+s\lambda \left(b+\frac{g\mu_BB}{2}\right)\right]^2+v^2k_z^2},
\end{align}
where $s,\lambda=\pm1$, $v$, $m$, $b$ are the model parameters, and $g$ is the $g$-factor \cite{SI}. We can see for $s\lambda=-1$, there is a competition between the $\sqrt{B}$ orbital term and the linear-$B$ Zeeman term. Because of the functional difference, there must be a critical $B$, at which the Zeeman term is equal to the orbital term, and the Landau bands $E_{\nu\pm\mp}$ become gapless. Above that critical magnetic field, the band bottom of $E_{\nu+-}$ becomes lower than the zeroth Landau band, leading to what we refer to the inversion of the Landau index, i.e.. the Landau band with a larger Landau index has a lower energy. There is no Landau index inversion for Schr\"odinger electrons, because their Landau bands also have the linear-$B$ dependence as the Zeeman energy does. The $\sqrt{B}$ orbital term in Eq. (2) is from the Dirac fermion nature of the model, thus the Landau index inversion gives a signature of Dirac fermions. Due to the small carrier density and large $g$-factor of ZrTe$_5$, it is easier to enter the quantum limit and have the Landau index inversion.

\begin{figure}[htbp]
\centering
\includegraphics[width=\columnwidth]{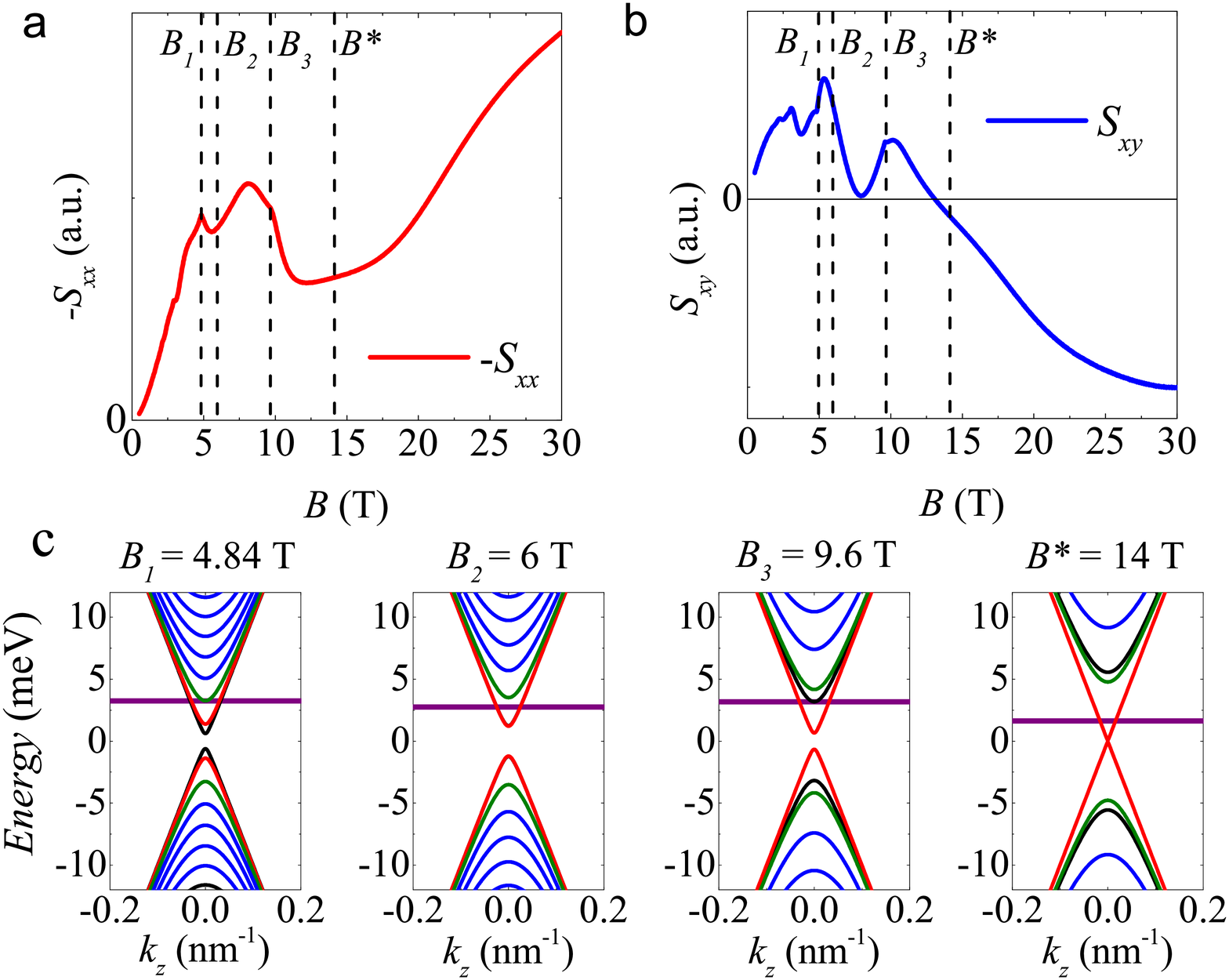}
\caption{ The calculated (a) $-S_{xx}$ and (b) $S_{xy}$ as functions of the magnetic field. (c) The Landau bands at several magnetic fields. The black, red, and green lines are for the 0th, 1st ($E_{1\pm\mp}$), and 2nd ($E_{2\pm\mp}$) Landau bands, respectively. The horizontal lines represent the Fermi energy $E_{F}$. The parameters are $v =$ 0.1 eVnm, $b = 0.034$ eV, $m =$ 0.036 eV, and $g$ = 18.65. The carrier density $n$ = 2.3 $\times 10^{16}$ cm$^{-3}$.}\label{fig.4}
\end{figure}

Our numerical results in Fig. 4 confirm the above mechanism. At $B_{1}$ $=$ 4.84 T, the Fermi energy begins to cut only the lowest two Landau bands and thermoelectric tensor undergoes quantum oscillations as higher Landau bands are depopulated. With increasing field, the bottom of the 0th Landau band increases, while the 1st Landau band decreases. At about $B_{2}$ $=$ 6 T, the 1st Landau band almost coincides with the 0th one. Above $B_2$, the 1st Landau band turns to have the lower energy than the 0th Landau band, i.e., there is a Landau index inversion. At $B_{3}$ $=$ 9.6 T, the Fermi energy begins to cut only the 1st Landau band and the anomalous peaks emerge in both $-S_{xx}$ and $S_{xy}$. Further increasing the magnetic field, the bottom of the 1st Landau band continues to decrease and, as shown in Fig. 4c, the system becomes gapless again at about $B^{\ast}$ $=$ 14 T.  The carrier density, instead of the Fermi energy, is fixed in the quantum limit at low temperatures \cite{ZhangCL19}. In this case, the Fermi energy shows a dramatic reduction when the Landau bands become gapless \cite{SI}. Therefore, the density of states and thermopower $-S_{xx}$ show a large dip. Moreover, the two terms in the Nernst signal $S_{xy}=\rho_{xx}\alpha_{xy}-\rho_{yx}\alpha_{xx}$ ($ \alpha_{\mu\mu'}=({\pi^2k_B^2T}/{3e}) \left.({\partial\sigma_{\mu\mu'}} /{\partial\epsilon})\right|_{\epsilon=E_F}$) have opposite signs. They compete with each other and have the same absolute value when the Landau bands are gapless, resulting in the sign reversal of $S_{xy}$. A recent observation of a cusplike feature in the magneto-infrared spectrum around 17 T also agrees well with our mechanism of Landau index inversion \cite{ChenZG}. Above $B^{\ast}$, the lowest Laudau bands reopen a gap, which leads to the growth in the thermopower. For simplicity, our calculation ignores the localization effects, so the calculated $-S_{xx}$ only has a dip near $B^{\ast}$ while the experimental $-S_{xx}(B^{\ast}) $ converges to zero. Nevertheless, according to our model, the large dip in $-S_{xx}$ and sign reversal in $S_{xy}$ are important characteristics of 3D Dirac fermions. In this sense, such high-field anomalous thermoelectric properties provide an effective experimental probe to 3D Dirac fermions.

\emph{Acknowledgments} -- This work was supported by the Natural Science Foundation of China (Grants No. 11474289, No. 11474005, No. U1432251, No. 11504378); Youth Innovation Promotion Association CAS (grant no. 2018486); the Innovative Program of Development Foundation of Hefei Center for Physical Science and Technology (grant no. 2017FXCX001); the Scientific Instrument Developing Project of the Chinese Academy of Sciences (grant no.YJKYYQ20180059) and the program of Users with Excellence, the Hefei Science Center of CAS. H. Z. Lu was supported by the Guangdong Innovative and Entrepreneurial Research Team Program (2016ZT06D348), the National Key R \& D Program
(2016YFA0301700), the National Natural Science
Foundation of China (11574127), and the Science, Technology and Innovation Commission
of Shenzhen Municipality (ZDSYS20170303165926217,
JCYJ20170412152620376).

\end{document}